\documentclass[prb,showpacs,twocolumn]{revtex4}

\usepackage{amsmath,amssymb,epsfig,epsf}

\begin{document}

\title
{Origin of ``hot-spots'' in the pseudogap regime of Nd$_{1.85}$Ce$_{0.15}$CuO$_{4}$: 
LDA+DMFT+$\Sigma_{\bf k}$ study}
\author{E.E. Kokorina$^1$, E.Z. Kuchinskii$^1$, I.A. Nekrasov$^1$, 
Z.V. Pchelkina$^2$, M.V. Sadovskii$^1$\\
A. Sekiyama$^3$, S. Suga$^3$, M. Tsunekawa$^3$}

\affiliation
{$^1$Institute for Electrophysics, Russian Academy of Sciences,
Ekaterinburg, 620016, Russia\\
$^2$Institute for Metal Physics, Russian Academy of Sciences, 
Ekaterinburg, 620219, Russia \\
$^3$Graduate School of Engineering Science, Osaka University, Toyonaka,
Osaka 560-8531, Japan}

\begin{abstract}


Material specific electronic band structure of the electron-doped high-T$_c$ cuprate
Nd$_{1.85}$Ce$_{0.15}$CuO$_4$ (NCCO) is calculated within the pseudo gap regime,
using the recently developed generalized LDA+DMFT+$\Sigma_{\bf k}$ scheme.
LDA/DFT (density functional theory within local density approximation) provides 
model parameters (hopping integral values, local Coulomb interaction strength) 
for the one-band Hubbard model, which is solved by DMFT (dynamical mean-field 
theory). To take into account pseudogap fluctuations LDA+DMFT is supplied with
``external'' ${\bf k}$-dependent self-energy $\Sigma_{\bf k}$, which
describes interaction of correlated conducting electrons with
non-local Heisenberg-like antiferromagnetic (AFM) spin fluctuations responsible for pseudo gap formation.
Within this LDA+DMFT+$\Sigma_{\bf k}$ approach we demonstrate the formation of
pronounced ``hot-spots'' on the Fermi surface (FS) map in NCCO, opposite to
our recent calculations for Bi$_2$Sr$_2$CaCu$_2$O$_{8-\delta}$ (Bi2212), which
have produced rather extended region of FS ``destruction''.
There are several physical reasons for this fact:
(i) the ``hot-spots'' in NCCO are located closer to Brillouin zone center;
(ii) correlation length of AFM fluctuations $\xi$ is larger for NCCO;
(iii) pseudogap potential $\Delta$ is stronger, than in Bi2212.
Comparison of our theoretical data with recent bulk sensitive
high-energy angle-resolved photoemission (ARPES) data for NCCO provides
good semiquantitative agreement.
Based on that comparison alternative explanation of the van-Hove singularity
at -0.3 eV is proposed.
Optical conductivity both for Bi2212 and NCCO is also calculated within
LDA+DMFT+$\Sigma_{\bf k}$ and compared with experimental results, demonstrating
satisfactory agreement.

\normalsize

\end{abstract}


\maketitle

\newpage

\section{Introduction}

There is a good reason to think that proper description of pseudogap regime
is the avenue of approach to the physical nature of high-$T_c$ superconductivity~\cite{psgap}.
Angle resolved photoemission spectroscopy (ARPES) has been coming along this
way very well in recent years. One of the test compounds for ARPES  
is the hole doped Bi$_2$Sr$_2$CaCu$_2$O$_{8-\delta}$ (Bi2212) system.
Another example is the electron doped high-T$_c$  
Nd$_{2-x}$Ce$_x$CuO$_4$ (NCCO). To this end there are plenty of experimental 
ARPES data on Bi2212 and NCCO (see the review~\cite{Bi2212}). Fermi surface 
(FS) maps, quasiparticle band dispersions and even self-energy lineshapes 
within mapping on some models are reliably extracted from modern 
ARPES data~\cite{Bi2212}.

There are several interesting physical phenomena associated with pseudo gap regime 
(in the normal underdoped phase): partial FS ``destruction'' and folding of band 
dispersions (shadow bands) for both compounds Bi2212 and NCCO \cite{Bi2212}. 
Despite evident similarities of experimental observations for these two systems 
there is one striking distinction. The FS of Bi2212 has so called Fermi ``arcs'' 
around ($\pi$/2,$\pi$/2) point (looking like a part of noninteracting FS), 
while towards BZ borders sharply defined FS just vanishes. In its turn NCCO 
also possesses slightly degraded Fermi ``arcs'', but in the vicinity of BZ borders noninteracting
FS is almost restored. In between there are well known ``hot-spots'' --- 
areas of FS ``destruction'' around the points where FS intersects 
umklapp BZ border. These ``hot-spots'' are not 
observed so obviously for Bi2212. The aim of the present paper is to show the 
origin of this NCCO ``hot-spots'' behavior.

Both systems under consideration are usually treated as Mott insulators at 
moderate doping or, in other words, as strongly correlated metals.
Modern technique to solve the Hubbard model is dynamical mean-field theory (DMFT),
which is exact in infinite dimensions \cite{georges96}.
However, the quasi two-dimensional nature of high-T$_c$ compounds is well known.
To overcome the local nature of the DMFT approximation, we have proposed 
recently a semiphenomenological DMFT+$\Sigma_{\bf k}$ computational scheme \cite{jtl,cm05,FNT}.
Here an additional self-energy $\Sigma_{\bf k}$ describes non-local correlations 
induced by (quasi) static short-ranged collective Heisenberg-like 
antiferromagnetic (AFM) spin (pseudogap) fluctuations \cite{Sch,KS}.
Assuming additive form of the self-energy within DMFT+$\Sigma_{\bf k}$ approach
one can preserve conventional DMFT self-consistent set of equations.
To take into account material specific properties of both Bi2212 and NCCO
we perform first principle one-electron density functional theory calculations within
local densty approximation (DFT/LDA) \cite{DFT_LDA}.
Then LDA results are incorporated into DMFT+$\Sigma_{\bf k}$ in accord with
LDA+DMFT ideology \cite{psik}.
To solve the effective single impurity problem of the DMFT we
employ here the reliable numerical renormalization group approach (NRG) 
\cite{NRG,BPH}. Such combined LDA+DMFT+$\Sigma_{\bf k}$ scheme by construction is particularly
suitable to describe electronic  properties of real high-T$_c$
materials at finite doping in the normal state.

The DMFT+$\Sigma_{\bf k}$ approach was extensively applied during the last years to
describe formation of pseudogap within strongly correlated metallic regime of 
single-band Hubbard model on the square lattice \cite{jtl,cm05,FNT}.
We have also generalized DMFT+$\Sigma_{\bf k}$ approach for the account of
static disorder effects \cite{FNT}. Later we have derived DMFT+$\Sigma_{\bf k}$ 
approach to calculate two-particle properties (such as optical conductivity)~\cite{SkOpt}.
Recently DMFT+$\Sigma_{\bf k}$ was used by us to analyze the general
problem of metal -- insulator transition in strongly disordered and strongly
correlated systems~\cite{HubDis}. 

LDA+DMFT+$\Sigma_{\bf k}$ scheme was already used to describe pseudogap regime 
in ``realistic'' calculations for Bi2212~\cite{skBi2212}. In the present paper 
we apply this approach to NCCO aiming to describe the characteristic differences
of its electronic structure as compared to Bi2212. 

The paper is organized as follows.
In section \ref{method} we present a short introduction into an
$ab~initio$ self-consistent generalized combined LDA+DMFT+$\Sigma_{\bf k}$ 
scheme and its extension for two particle properties (optical conductivity).
Section \ref{lda} contains Bi2212 and NCCO material specific information:
LDA calculated band structure, Fermi surfaces and details on some model 
parameters calculations.  Results and discussion of LDA+DMFT+$\Sigma_{\bf k}$ 
calculations for Bi2212 and NCCO and comparisons with experimental data
are presented in section \ref{results}. Section \ref{concl} concludes our 
manuscript with the summary and discussion of some of the remaining problems.

\section{Computational method}
\label{method}

\begin{figure}
\includegraphics[clip=true,width=1.\columnwidth]{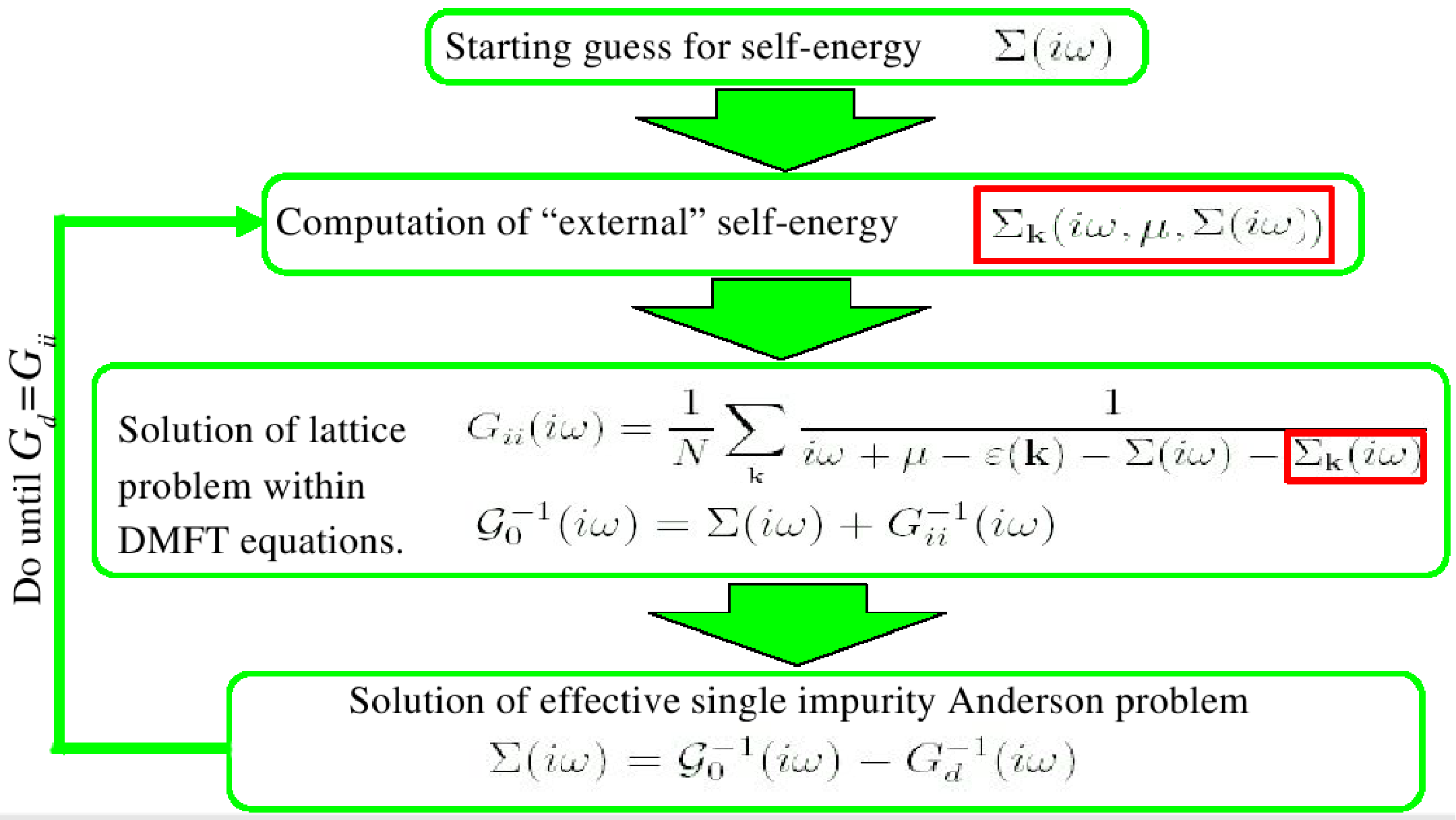}
\caption{Flow diagram of DMFT+$\Sigma_{\bf k}$ selfconsistent loop. 
$ii$ corresponds to lattice problem and
$d$ -- impurity problem variables.}
\label{scheme}
\end{figure}

To introduce spatial length scale (nonlocal correlations) into conventional
DMFT method \cite{georges96} we have
recently proposed the generalized DMFT+$\Sigma_{\bf k}$ approach \cite{jtl,cm05,FNT},
with computational scheme, shown in Fig.~\ref{scheme}, which contains the flow 
diagram of self consistent DMFT+$\Sigma_{\bf k}$ set of equations.
First we guess some initial local (DMFT) electron self-energy $\Sigma(i\omega)$.
Second we compute by any available technique (for the chosen model) ${\bf k}$-dependent ``external''
self-energy $\Sigma_{\bf k}(i\omega)$ which can be in general case
a functional of $\Sigma(i\omega)$.
Then neglecting interference effects between the self-energies
(which in fact is the major assumption of our approach)
we can set up and solve the lattice problem of DMFT (step 3 in Fig.~\ref{scheme}).
Then at step 4 we define effective Anderson single impurity problem which is to be solved
by any ``impurity solver'' to close DMFT+$\Sigma_{\bf k}$ equations.

The additive form of self-energy (at step 3 in Fig.~\ref{scheme})
is in fact an advantage of our DMFT+$\Sigma_{\bf k}$ approach~\cite{jtl,cm05,FNT}.
It allows one to preserve the set of self-consistent equations of standart DMFT
\cite{georges96}.
However there are two distinctions from conventional DMFT.
During each DMFT iteration we recalculate corresponding  {\bf {k}}-dependent self-energy
$\Sigma_{\bf k}(\mu,\omega,[\Sigma(\omega)])$ within some (approximate) scheme,
e.g. taking into account interactions with collective modes or order parameter
fluctuations, and the local Green function $G_{ii}(i\omega)$ is ``dressed'' 
by $\Sigma_{\bf k}$ at each step. When input and output Green's functions 
(or self-energies) converge to each other (with prescribed accuracy)
we consider the obtained solution to be self-consistent.
Physically it corresponds to the account of some ``external'' (e.g. pseudogap) 
fluctuations, characterized by an important length scale $\xi$,
into fermionic ``bath'' surrounding the effective Anderson impurity of the 
usual DMFT.

In the present work $\Sigma_{\bf k}(\omega)$ represents
interaction of correlated electron with antiferromagnetic (AFM) pseudogap 
fluctuations. To calculate $\Sigma_{\bf k}(\omega)$ for
the case of random field of pseudogap fluctuations 
(assumed to be (quasi) static and Gaussian, which is valid at high enough 
temperatures \cite{Sch,KS}) with the
dominant scattering momentum transfers of the order of characteristic vector 
${\bf Q}=(\pi/a,\pi/a)$ ($a$ - lattice parameter), typical for AFM fluctuations 
(``hot-spots'' model \cite{psgap}), the recursion procedure 
proposed in Refs.~\cite{MS79,Sch,KS} is used, with material specific 
generalizations as described in detail in Refs.~\cite{skBi2212}.

There are two important parameters characterizing pseudogap regime in our
scheme: the pseudogap energy scale (amplitude) $\Delta$ and spatial 
correlation length $\xi$ \cite{KS,skBi2212}. Actually we prefer to take $\Delta$ 
and $\xi$ determined somehow from experiments. However, we can also
use certain model estimates to calculate them microscopically \cite{cm05}. Both
approaches are used below.

To calculate optical 
conductivity we use our generalization of DMFT+$\Sigma_{\bf k}$ for calculation
of two -- particle properties (vertex parts) as described in detail in 
Ref.~\cite{SkOpt}, with material dependent parameters provided by
LDA+DMFT+$\Sigma_{\bf k}$ and vertex corrections due to pseudogap fluctuations
calculated using the recursion relations derived earlier in Ref.~\cite{SS02}

\section{LDA bands and FS of NCCO and Bi2212, effective model parameters}
\label{lda}

\begin{figure}
\includegraphics[clip=true,angle=270,width=0.75\columnwidth]{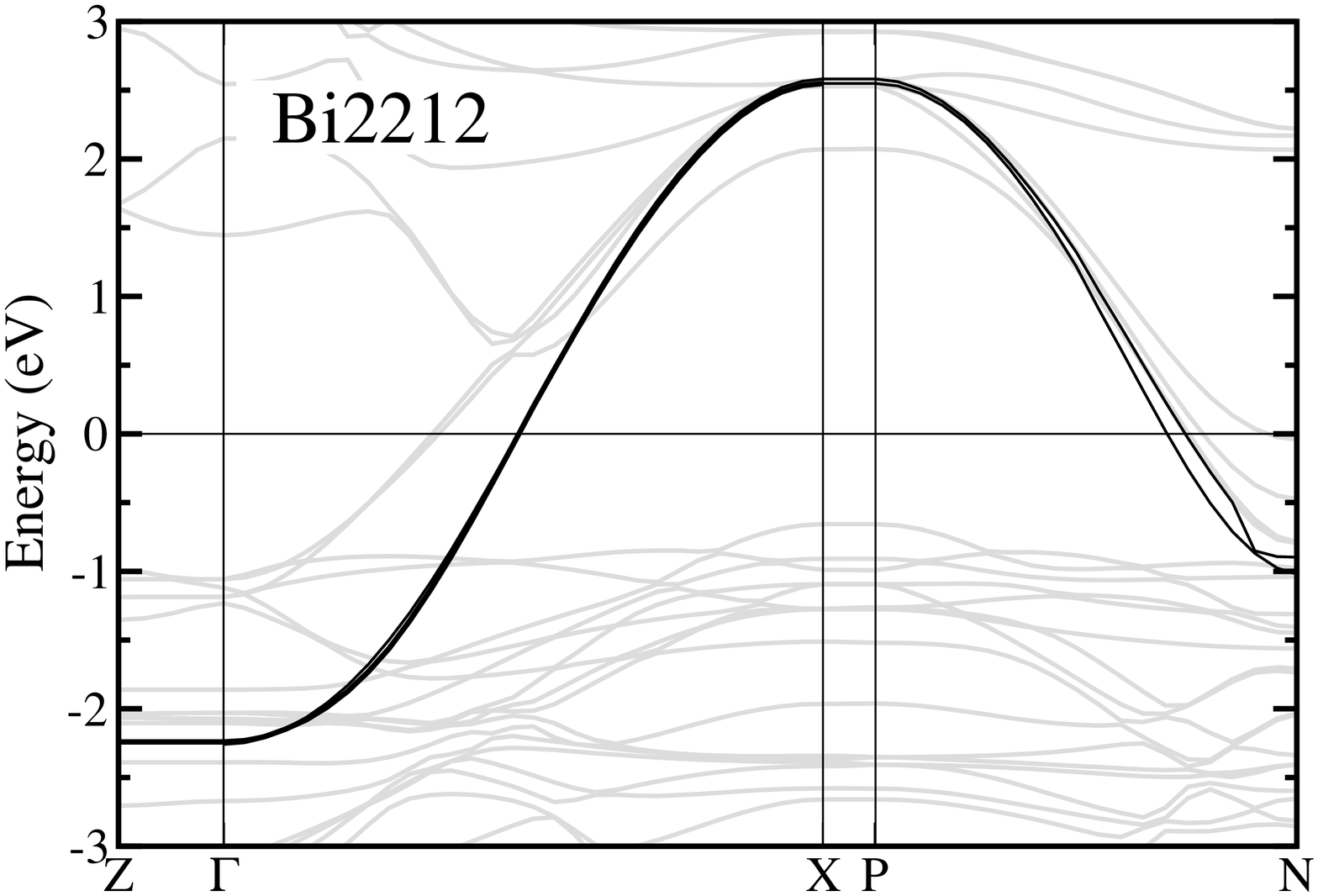}
\includegraphics[clip=true,angle=270,width=0.85\columnwidth]{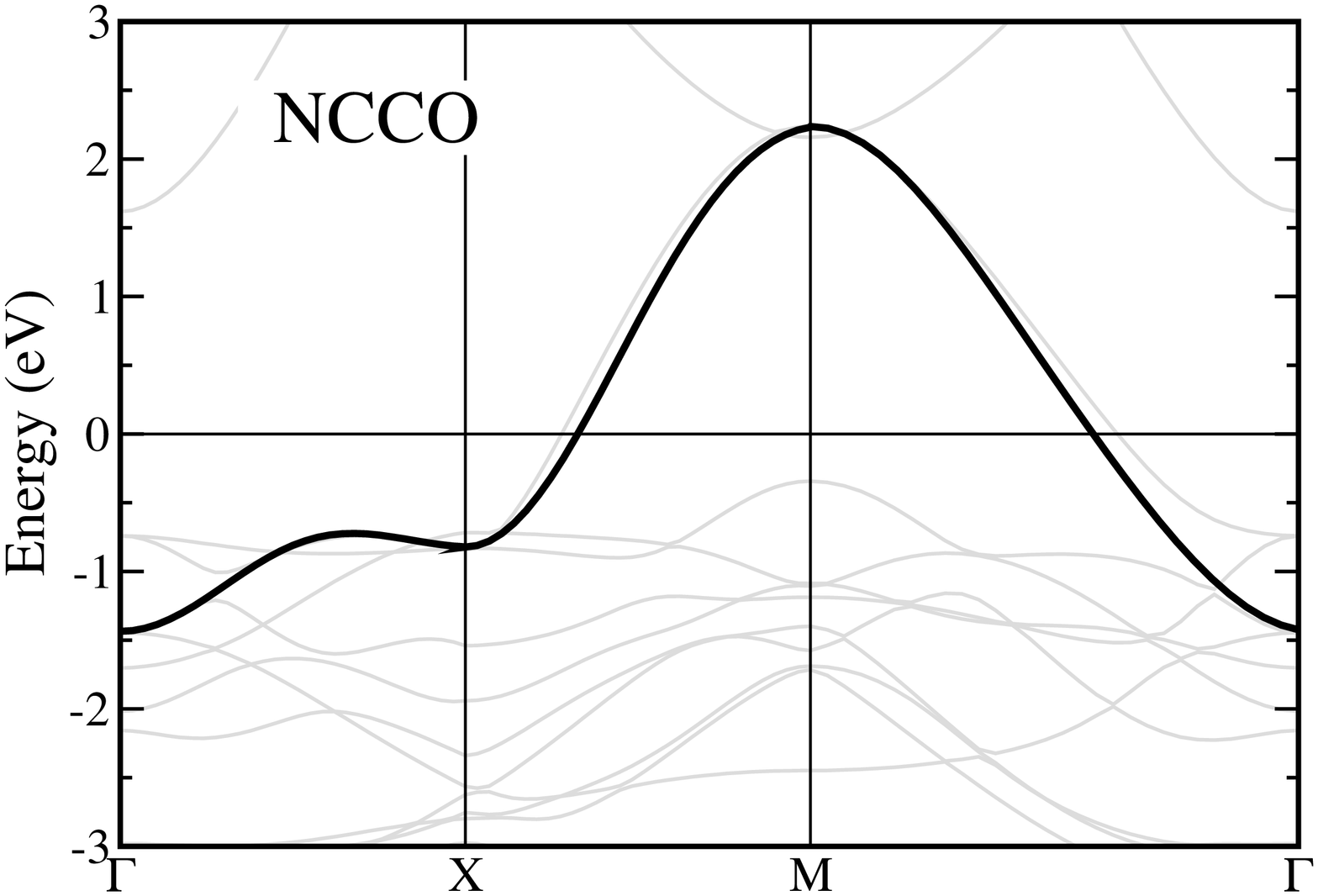}
\caption{LDA bands (gray lines) for Bi2212 (left) and NCCO (right) along BZ high symmetry directions.
For both panels solid black lines correspond to effective $x^2\!-\!y^2$ symmetry Wannier-like state dispersions.
Zero energy corresponds to Fermi level.}
\label{bands}
\end{figure}

As a first step of our LDA+DMFT+$\Sigma_{\bf k}$ hybrid scheme we perform LDA band structure calculations. For both compounds ideal tetragonal bcc crystal lattice with space symmetry group I4/mmm is reported (for Bi2212 see Ref.~\cite{Bi2212struct}, for NCCO Ref.~\cite{NCCOstruct}). Physically relevant structural motif for high-T$_c$
materials is CuO$_2$ plane. There are two CuO$_2$ planes displaced close to each other 
in the unit cell of Bi2212, while NCCO has just one.
We have done LDA calculations of electronic band structure within the linearized muffin-tin orbital (LMTO)
basis set~\cite{LMTO}. Results are presented as gray lines in Fig.~\ref{bands}:
left panel --- Bi2212 bands, right --- NCCO.
Our band structures agree well with previous works Ref.~\cite{BiLDA} and Ref.~\cite{NdLDA} for Bi and Nd compounds correspondingly.

To calculate hopping integral values for Bi system Wannier functions projecting method~\cite{Marzari} in the LMTO framework~\cite{Anisimov05} was applied. Hopping integrals of Nd compound were obtained by using the so
called NMTO method~\cite{NMTO} (see Table~\ref{param}). Values of hopping integrals computed by these two methods
agree well for the respective compounds~\cite{Korshunov04}.
In Fig.~\ref{bands} by black line represents the dispersion of effective
$x^2-y^2$ Wannier-like orbital which crosses the Fermi level and is most 
interesting physically. These dispersions correspond to 
hopping integral values (tight-binding parameters) given in Table~\ref{param}.

\begin{table}[b]
\label{param}
\caption {Calculated energetic model parameters for Bi2212 and NCCO (eV). First four Cu-Cu in plain hopping integrals
$t$, $t^{\prime}$, $t^{\prime\prime}$, $t^{\prime\prime\prime}$, interplain hopping value~$t_\perp$,
local Coulomb interaction~$U$ and pseudogap potential~$\Delta$.}
\begin{tabular}{| c | c | c | c | c | c | c | c |}
\hline
       &   $t$  & $t^{\prime}$  & $t^{\prime\prime}$ &  $t^{\prime\prime\prime}$ &  $t_\perp$ & $U$ & $\Delta$\\
\hline
Bi2212 & -0.627 & 0.133 & 0.061 & -0.015  & 0.083 & 1.51 & 0.21\\
\hline
NCCO   & -0.44  & 0.153 & 0.063 & -0.0096 & ---   & 1.1  & 0.36\\
\hline
\end{tabular}
\end{table}

\begin{figure}
\includegraphics[clip=true,width=0.6\columnwidth]{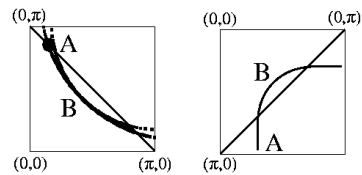}
\caption{LDA calculated Fermi surfaces for Bi2212 (left) and NCCO (right) in the quarter of the Brillouin zone.
Diagonal line corresponds to (AFM) umklapp scattering surface.}
\label{ldafs}
\end{figure}

Fig.~\ref{ldafs} contains noninteracting LDA Fermi surfaces (FS) in $(k_x,k_y)$ plane for
quarter of the first Brillouin zone (BZ) (left panel -- Bi, right panel -- Nd systems).
The shape of these FS is defined by the tight-binding parameters from Table~\ref{param}.
Diagonal line corresponds to the AFM folded BZ border. In the left panel for Bi2212 one can see two FS sheets.
It is caused by finite hopping between two neighboring CuO$_2$ layers --- the so called bilayer splitting.
The value $t_\perp$ is listed in the Table~\ref{param}. Simplest tight-binding expression for the bilayer splitting
derived by Andersen $et~al.$~\cite{Andersen95} is used in our 
calculations (for details see Ref.~\cite{skBi2212}).

It is important to note the ``hot-spots'' positions (intersections of FS with 
AFM umklapp surface) for both materials. For Bi2212 it is (0.47,2.66)$\pi/a$ and for NCCO (0.95,2.19)$\pi/a$.
Whereof one can see that for NCCO the ``hot-spots'' are located farther away from the BZ border than in
Bi2212. Let us remind that pseudogap fluctuations scatter electrons from the vicinity of
one ``hot-spot'' to the vicinity of another ``hot-spot'', i.e. by the scattering
vector of the order of ${\bf Q}$. The effective scattering area around the
``hot-spot'' is defined by inverse correlation length $\xi^{-1}$ of these
fluctuations.
Let us also remember that ($\pi/a$,0) point is surrounded by four BZ from different sides.
Consequently if the ``hot-spot'' is closer to ($\pi/a$,0) point and 
$\xi$ is small enough,  FS is ``destroyed'' in rather wide region
close to BZ border crossings. Thus one can expect for NCCO ``hot-spots'' to be 
observed more explicitly, while the part of the FS close to the BZ border crossings
be less affected by pseudogap fluctuations in contrast to Bi2212.
 
The values of local Coulomb interaction $U$ for $x^2-y^2$ orbital were obtained by constrained LDA
method~\cite{Gunnarsson88} (for values see Table~\ref{param}). These values are of the order of 2-3 $t$
for both systems. It is in fact quite a bit smaller than the values many people believe should be used
in a model calculations (usually about 4-6 $t$, see e.g. Ref.~\cite{Plakida}).
At the same time, our previous experience with constrained LDA computation 
shows that it gives reasonable estimates for the $U$ value in a number of 
other oxides~\cite{Nekrasov00}. However, to analyze further the 
influence of $U$ value on observable quantities we performed additional
LDA+DMFT+$\Sigma_{\bf k}$ computations for increased values of $U$. 
The short discussion of the results will be given in Sec.~\ref{concl}.
Values of $\Delta$ for both systems were calculated as proposed in 
Ref.~ \cite{cm05} (Table~\ref{param}).
(for more details see Appendix~\ref{app1}). Correlation length $\xi$ was
taken from experiments, i.e. $\xi\sim10a$
for Bi2212~\cite{psgap} and $\xi\sim50a$ for NCCO~\cite{NCCOxi}.

\section{NCCO vs. Bi2212 LDA+DMFT+$\Sigma_{\bf k}$ results and experimental data}
\label{results}

\subsection{Quasiparticle dispersions}
\label{qpdisp}

\begin{figure}
\includegraphics[clip=true,width=0.6\columnwidth]{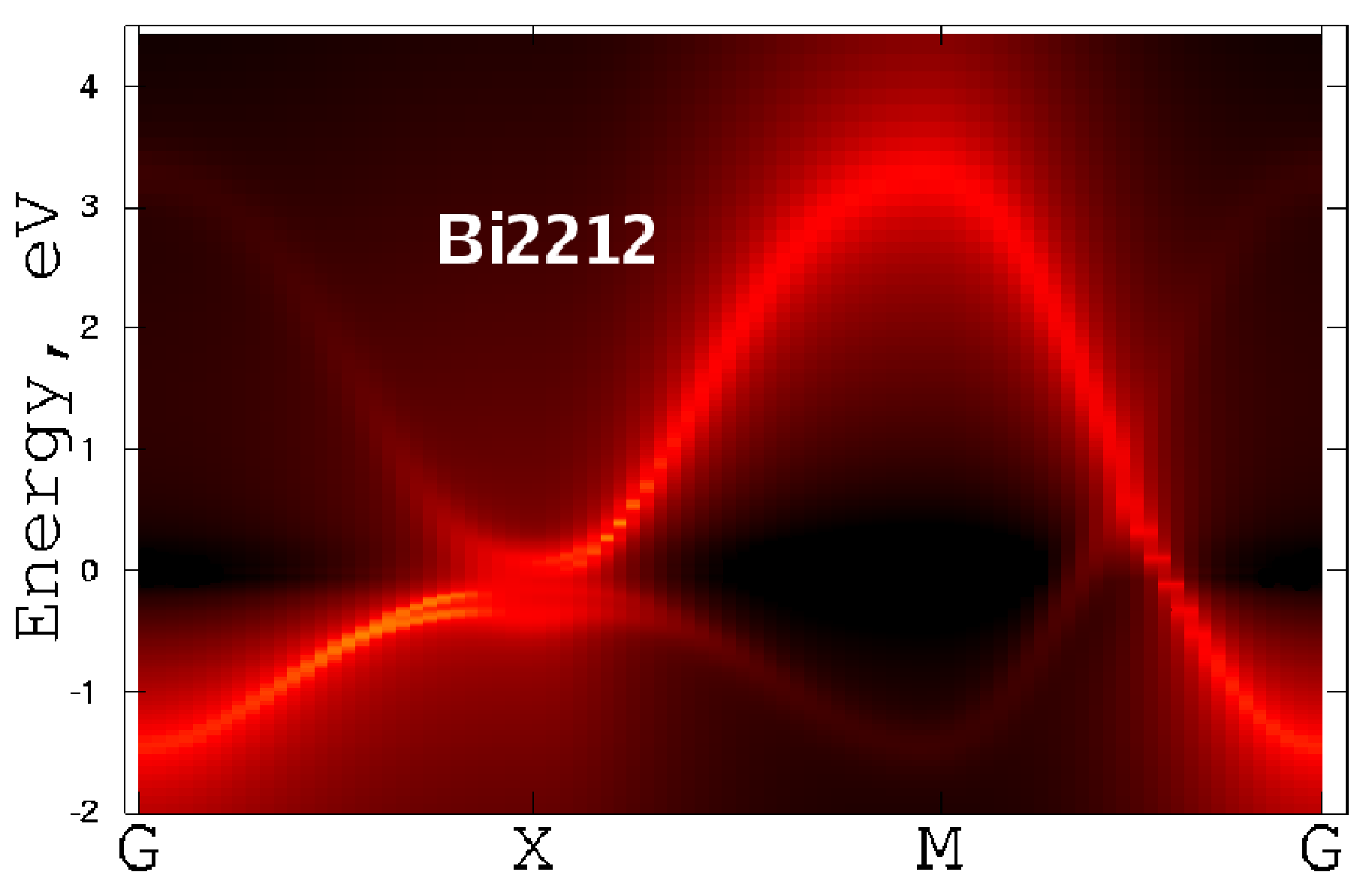}
\includegraphics[clip=true,width=0.6\columnwidth]{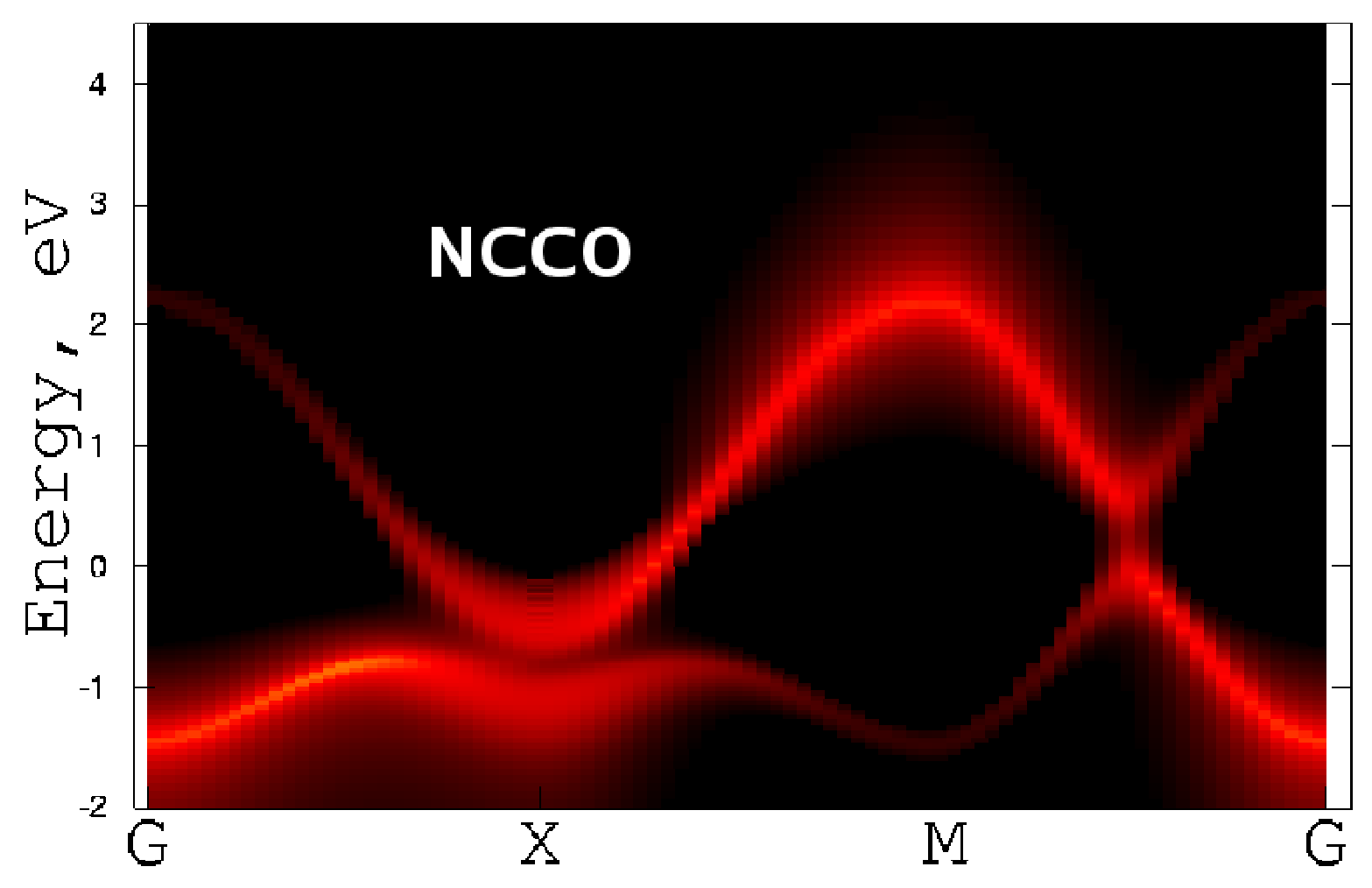}
\caption{LDA+DMFT+$\Sigma_{\bf k}$ spectral functions contour plot for Bi2212 (upper panel) and NCCO (lower panel)
along BZ high-symmetry directions of first Brillouin zone
$\Gamma(0,0)\!-\!\rm{X}(\pi,0)\!-\!\rm{M}(\pi,\pi)\!-\!\Gamma(0,0)$.
Zero energy corresponds to Fermi level.}
\label{sdtri}
\end{figure}

Finite temperatures and interactions lead in general to finite life-time
effects. Thus, instead of quasiparticle dispersions expressed by the usual 
dispersion curves (like, for example, in DFT/LDA) in Fig.~\ref{sdtri} we display 
contour plots of corresponding spectral functions $A(\omega,{\bf k})$:
\begin{equation}
A(\omega,{\bf k})=-\frac{1}{\pi}{\rm Im} G(\omega,{\bf k}),
\label{specf}
\end{equation}
where $G(\omega,{\bf k})$ is the retarded Green's function obtained via our
LDA+DMFT+$\Sigma_{\bf k}$ scheme (shown in Fig.~\ref{scheme}) with appropriate 
analytic continuation to real frequencies.

$Prima~facie$ both compounds Bi2212 (left panel of Fig.~\ref{sdtri}) and NCCO (right panel of Fig.~\ref{sdtri})
have similar quasiparticle bands. There are two bands in each case instead of just one in case of DFT/LDA.
Of these,  the most broad and intensive band predominantly follows the noninteracting DFT/LDA band (see Fig.~\ref{bands}).
The second band in our case is an AFM-like reflex (shadow band) of the latter one and has much weaker intensity.
This is a direct effect of self-energy $\Sigma_{\bf k}$ due to pseudogap 
fluctuations introduced into the conventional DMFT (for discussion see 
Ref.~\cite{modband}). 

As we discussed above, finite life-time (interaction) effects should be 
especially strong around the
($\pi/a$,0) point ($X$-point in Fig.~\ref{sdtri}).
This is clearly visible in both panels of Fig.~\ref{sdtri}.
However in Sec.~\ref{lda} we showed that Bi2212 ``hot-spot'' is much closer to the $X$-point.
Thus we see large quasiparticle band broadening (do not forget bilayer splitting effects)
on the Fermi level. For NCCO there is no sizeable broadening on the Fermi level close to $X$-point.
Both branches go below Fermi level at about -0.5 eV.
In NCCO the Fermi ``arc'' is much closer to umklapp surface, so pseudogap
effects are significant around ($\pi/2a$,$\pi/2a$) point (middle point of the $\Gamma-M$ direction)
in contrast to Bi2212.

\begin{figure}
\includegraphics[clip=true,angle=270,width=0.6\columnwidth]{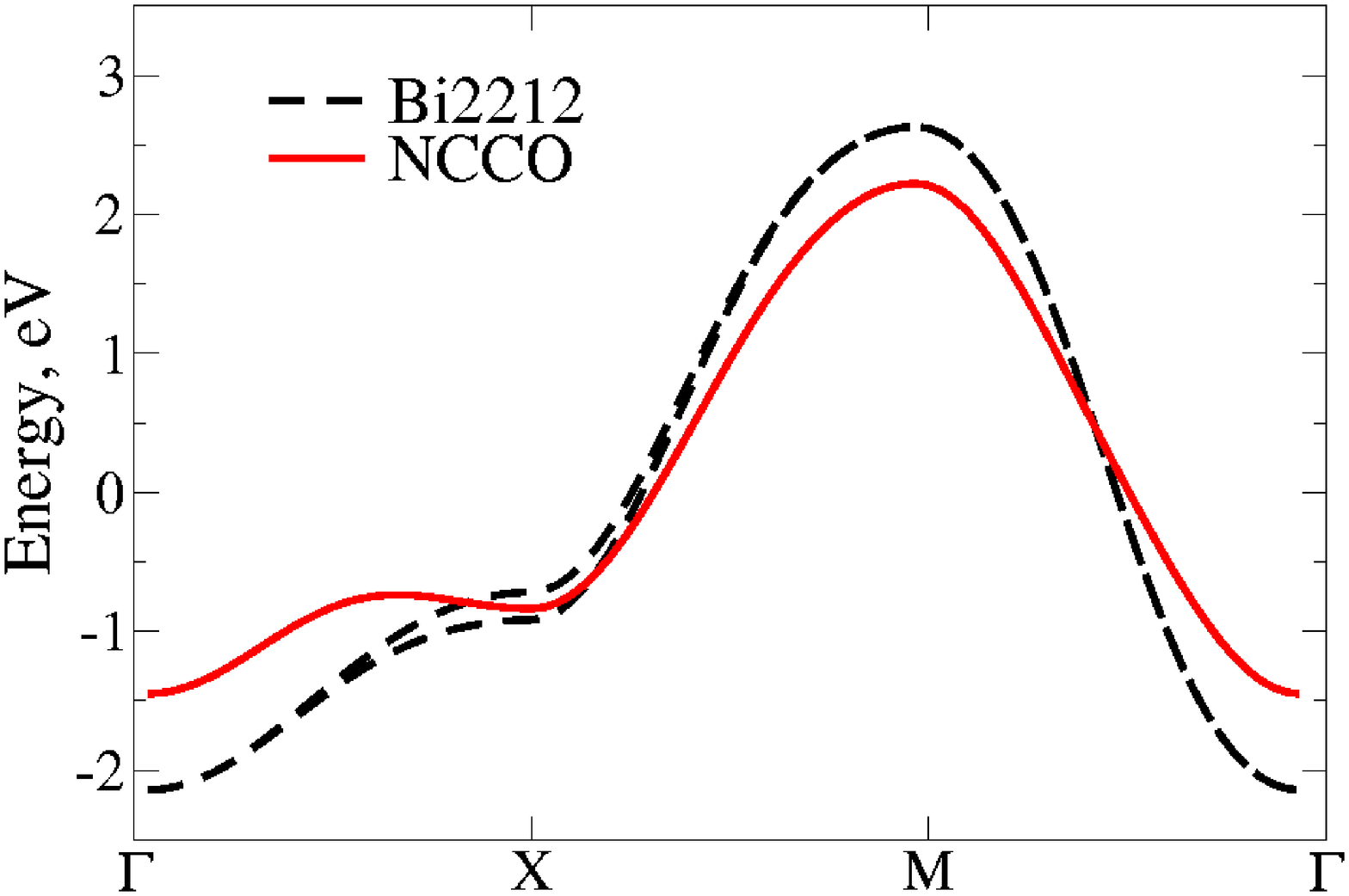}
\includegraphics[clip=true,angle=270,width=0.6\columnwidth]{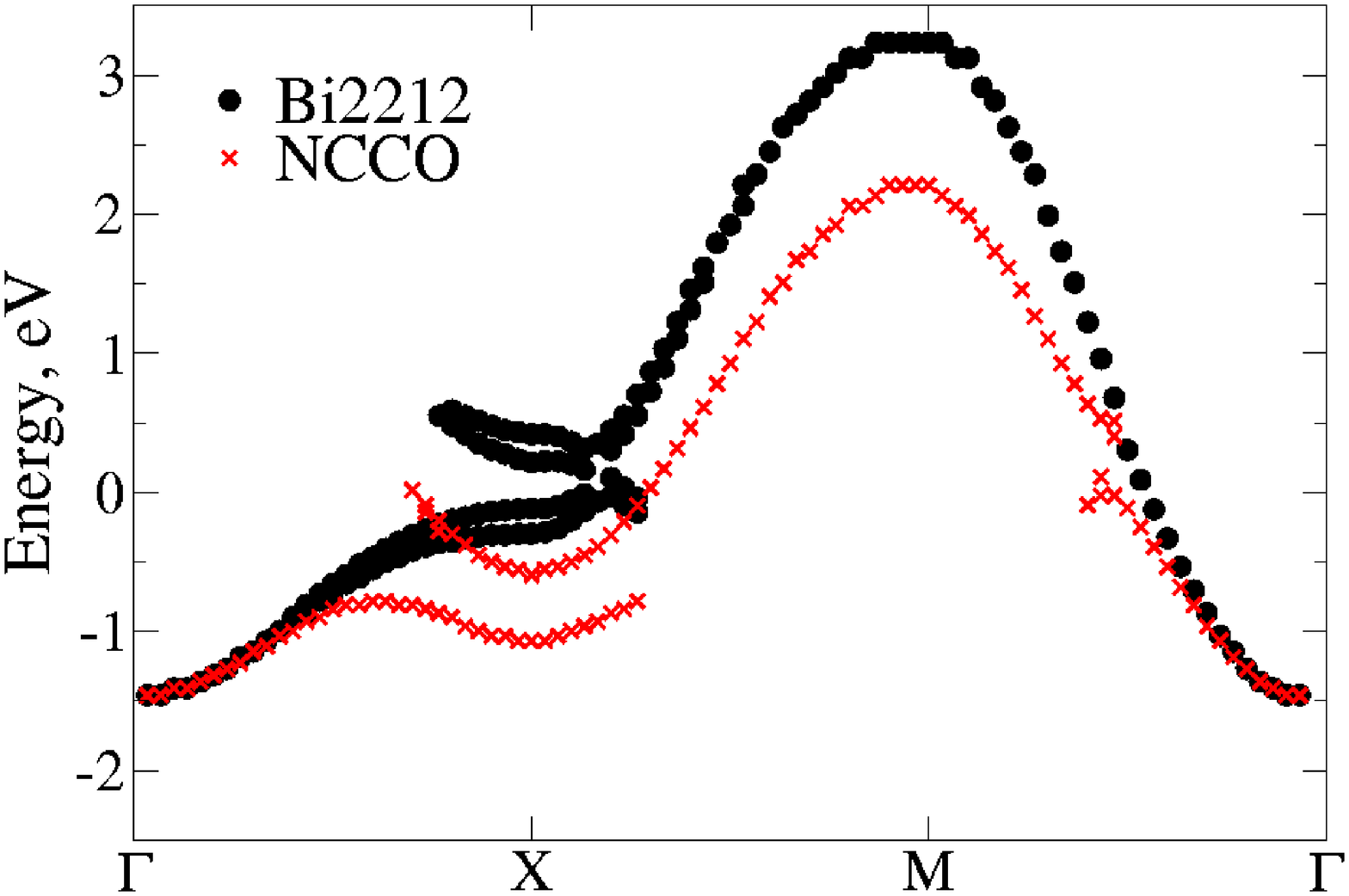}
\caption{Comparison of effective LDA $x^2-y^2$ bands (upper panel)
and LDA+DMFT+$\Sigma_{\bf k}$ quasiparticle dispersions (lower panel)
Bi2212 and NCCO along BZ high-symmetry directions.
Zero energy corresponds to Fermi level.}
\label{compband}
\end{figure}

Fig.~\ref{compband} shows changes from effective $x^2-y^2$ LDA bands (left side)
to LDA+DMFT+$\Sigma_{\bf k}$ quasipartical bands (right side) for both Bi2212 and NCCO.
Quasiparticle bands on the right panel of Fig.~\ref{compband} 
represent the positions of the maxima of spectral functions shown in Fig.~\ref{sdtri}.
In Bi2212 the shadow band and the quasiparticle band
intersect each other in the ``hot-spot'' close to the $X$-point.
While in NCCO there is no such intersection but the shadow and quasiparticle bands are quite parallel
around $X$-point. Close to the ($\pi/2a$,$\pi/2a$) point we observe 
a kind of precursor of the dielectric AFM gap. Nothing of that sort is observed 
for Bi2212. Also one should mention that calculated shadow band in Bi2212 is 
actually an order of magnitude less intensive than in NCCO.

\subsection{Spectral functions}

\begin{figure}
\includegraphics[clip=true,angle=270,width=0.45\columnwidth]{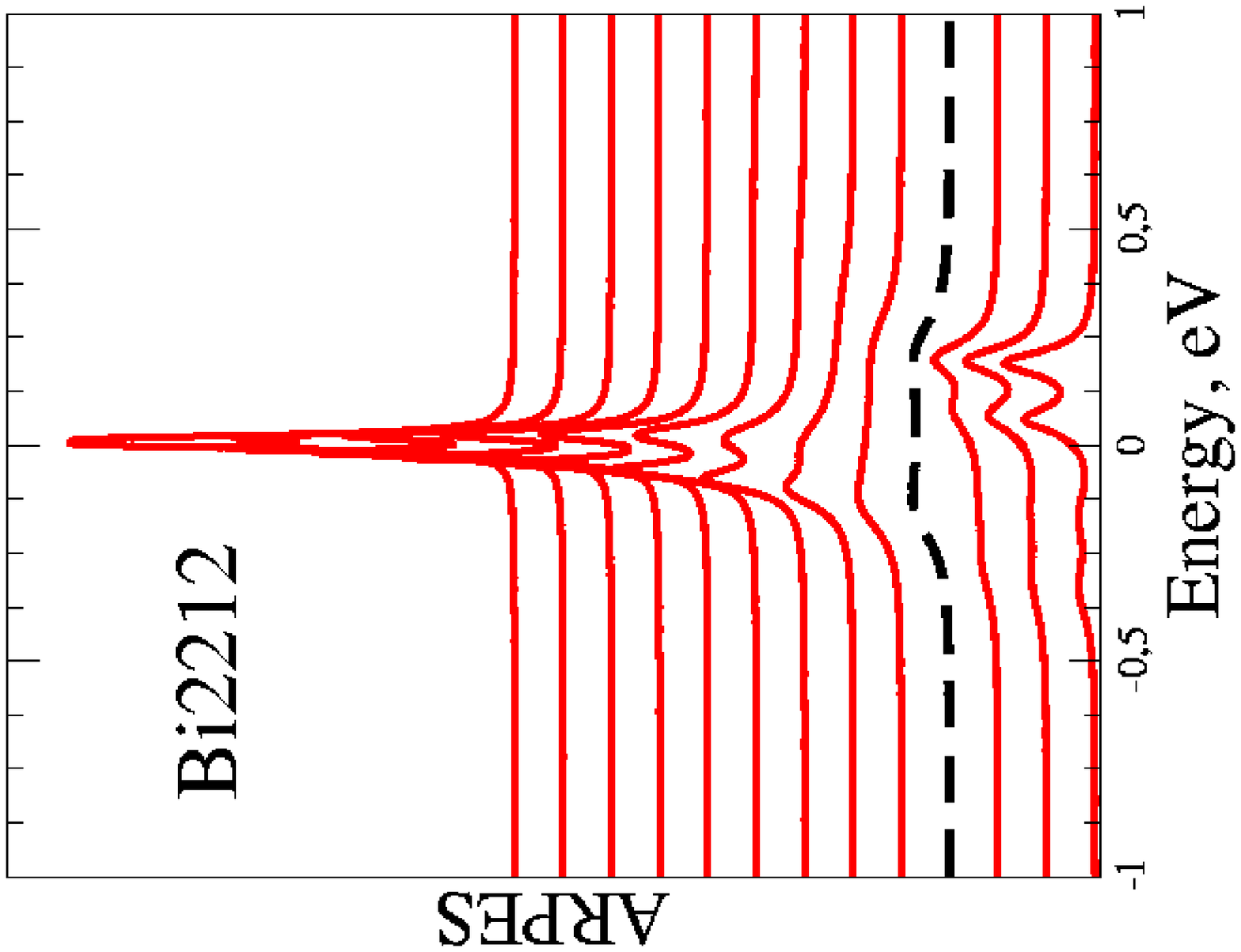}
\includegraphics[clip=true,angle=270,width=0.45\columnwidth]{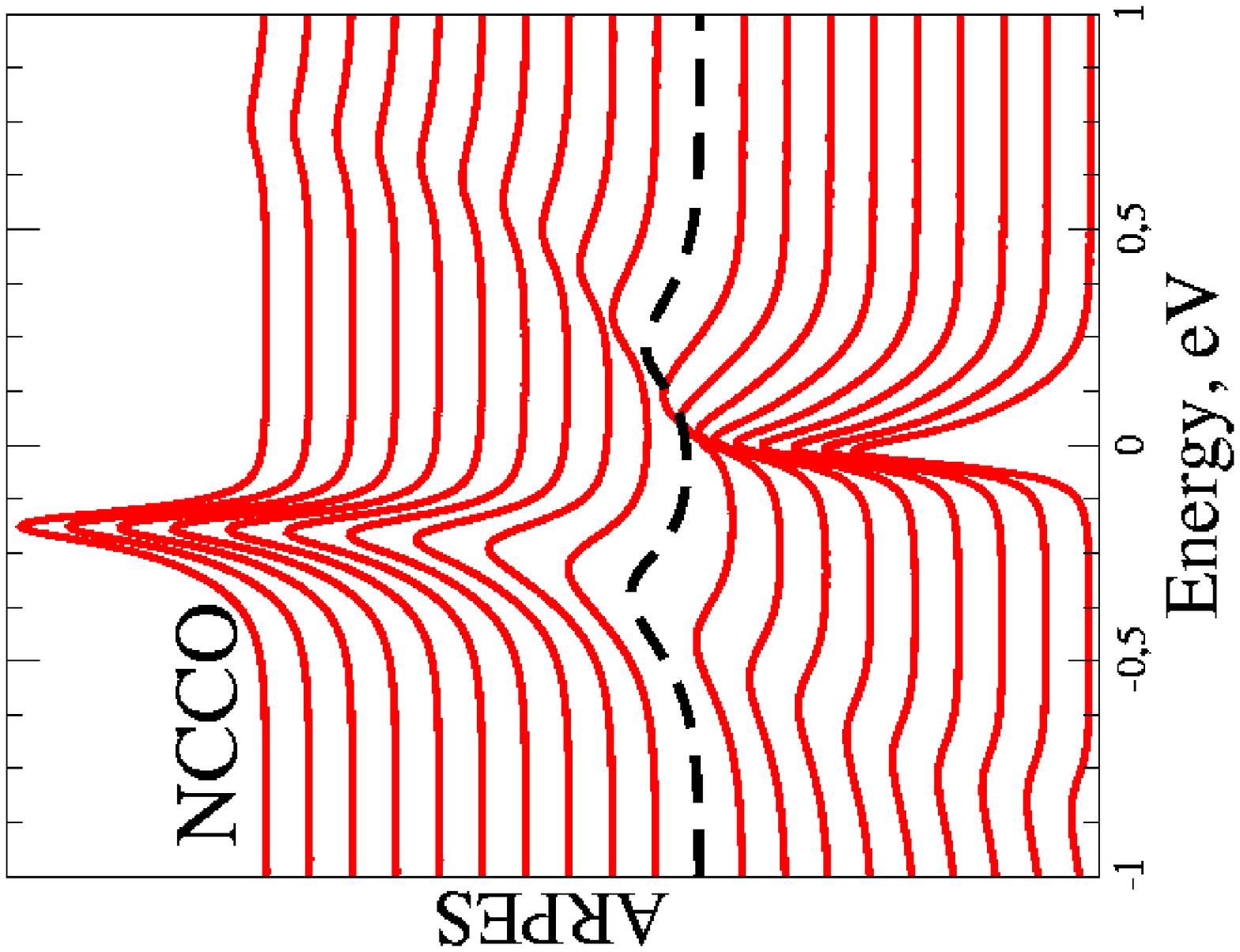}
\caption{LDA+DMFT+$\Sigma_{\bf k}$ spectral functions for Bi2212 (upper panel) and NCCO (lower panel)
along of noninteracting FS in 1/8 of BZ. Dashed-black line corresponds to ``hot-spot''.}
\label{sdfs}
\end{figure}

Fig.~\ref{sdfs} displays LDA+DMFT+$\Sigma_{\bf k}$ spectral functions (\ref{specf}) along 1/8 of noninteracting FS from the nodal point (top curve) to the antinodal one (bottom curve)  (A and B points in Fig.~\ref{ldafs}, correspondingly).
Data for Bi2212 is given in left panel, NCCO --- right panel of Fig.~\ref{sdfs}. For both compounds  antinodal quasiparticles are well-defined --- sharp peak close to the Fermi level. Going to the nodal point quasiparticle damping
grows and peak shifts to higher binding energies. This behavior is confirmed
by experiments Refs.~\cite{Armitage02,Kaminski05} (for brief comparison 
with experiment see Ref.~\cite{Nekrasov07}). Again there are some differences 
between these two compounds. As we said before ``hot-spots'' for NCCO are 
closer to the BZ center. In Fig.~\ref{sdfs} one can see it from the position 
of the dashed-black line which corresponds to the ``hot-spots'' ${\bf k}$-point.  
Thus another explanation of the peaks can be given. Namely, for Bi2212 nodal quasiparticles are formed by low energy edge of pseudogap. While for NCCO they are formed by higher energy pseudogap edge. Also in NCCO there is obviously
no bilayer splitting effects seen for Bi2212 (left panel of Fig.~\ref{sdfs}). 

\subsection{Comparison with ARPES data}
 
\begin{figure}

\includegraphics[clip=true,width=\columnwidth]{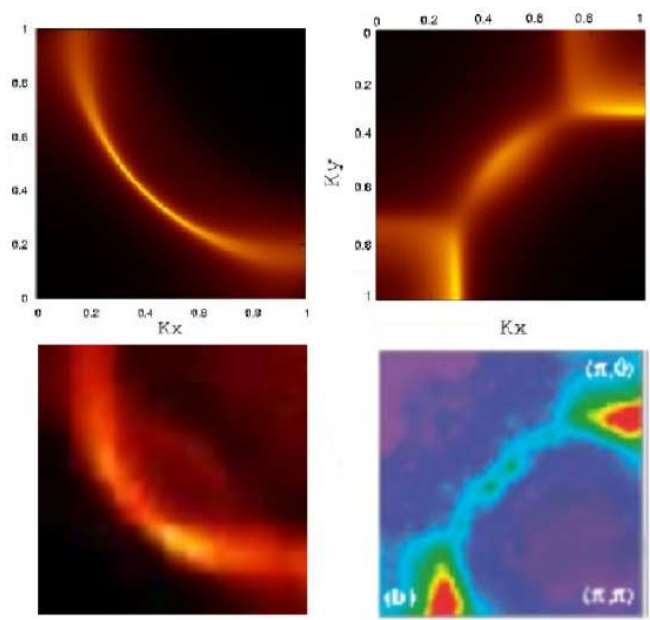}
\caption{LDA+DMFT+$\Sigma_{\bf k}$ Fermi surfaces for Bi2212 (upper left panel) and NCCO (upper right panel)
1/4 of BZ ($k_x,k_y$ in units of $\pi/a$). Experimental FS for Bi2212 (lower left panel, Ref.~\protect\cite{Borisenko00}) and NCCO (lower right panel, Ref.~\protect\cite{Armitage02}).}
\label{sk_fs}
\end{figure}

In Fig.~\ref{sk_fs} LDA+DMFT+$\Sigma_{\bf k}$ FS maps on the quarter of BZ
for Bi2212 (upper left) and NCCO (upper right) are presented.
The upper parts of Fig.~\ref{sk_fs} are just a contour plot of the spectral functions from Fig.~\ref{sdfs}.
Above mentioned significant FS ``destruction'' because of pseudogap fluctuations close to the borders of BZ
is clearly seen for Bi2212. Quite the contrary NCCO FS is almost restored in the vicinities
of the BZ border. Vice versa Fermi ``arc'' of Bi2212 is quite sharp while for NCCO is rather degraded.
That is afresh the consequence of the ``hot-spots'' to be closer to BZ center for NCCO.
Slightly bigger value of pseudogap potential $\Delta$ also works towads Fermi ``arc'' smearing in NCCO.
It is significant to say that shadow FS are come to hand. Shadow FS is found to be more
intensive for NCCO.

Qualitatively speaking very similar FS shapes are observed experimentally for both Bi Ref.~\cite{Borisenko00}
and Nd Ref.~\cite{Armitage02} compounds (lower parts of Fig.~\ref{sk_fs}).
To our opinion such FS maps have material specific origin. 
Namely LDA calculated FS of NCCO has more curvature (left panel of Fig.~\ref{ldafs})
and intersects the BZ boundary away from the ($\pi/a$,0) thus remaining nearly noninteracting one.
While Bi2212 FS comes to BZ border much closer to the ($\pi/a$,0) point (right panel of Fig.~\ref{ldafs}).
Because of that in Bi2212 ``hot-spots'' are not seen. They are spread by strong pseudogap 
scattering processes near ($\pi/a$,0) point.
Also bigger correlation length for NCCO is favorable for more evident ``hot-spots''.

\begin{figure}
\includegraphics[clip=true,width=0.8\columnwidth]{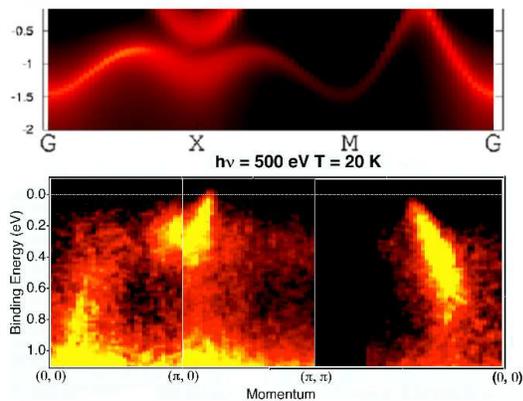}
\caption{Comparison of LDA+DMFT+$\Sigma_{\bf k}$ spectral functions 
(upper panel) for NCCO along BZ high-symmetry directions with experimental 
ARPES (Ref.~\protect\cite{Tsunekawa}) (lower panel).}
\label{ncexp}
\end{figure}

In Fig.~\ref{ncexp} we present LDA+DMFT+$\Sigma_{\bf k}$ data in comparison with recent
high-energy bulk sensitive angle-resolved photoemission data of Nd$_{1.85}$Ce$_{0.15}$CuO$_4$.
For the details of experiment see Ref.~\cite{Tsunekawa}.
Lower panel of the  Fig.~\ref{ncexp} shows intensity plots along the high symmetry lines
for NCCO obtained by the high-h$\nu$ ARPES.
Upper panel of Fig.~\ref{ncexp} is part of the Fig.~\ref{sdtri}.
Here we should say that to get better agreement with experiment theoretical Fermi level
is changed by 0.2 eV.

Indeed we see quite a good agreement of LDA+DMFT+$\Sigma_{\bf k}$ and experimental data.
For the $M-\Gamma$ direction there is not very much going on. Basically we see both in theory and experiment
very intensive quasiparticle band. For the $M-\Gamma$ direction low intensive shadow band is not resolved
in the experiment.

More interesting situation is observed for $\Gamma-X-M$ directions.
At $\Gamma$-point there is band in the experiment starting at about -1.2 eV.
It is rather intensive and goes up in energy. Suddenly there is almost zero intensity
at about -0.3 eV. Then in the vicinity of the $X$-point intensity rise up again.
In the $X-M$ direction around -0.3 eV on the right side of $X$-point there is also quite intensive
region. At a first glance one can think that it is the same band with matrix element effects
governing intensity. However looking at the right panel Fig.~\ref{compband} (see corresponding discussion
in Sec.~\ref{qpdisp}) one can realize that this low intensity region is the forbidden gap between
shadow and quasiparticle bands. The ``horseshoe'' around $X$-point is formed by the shadow band on the left and
the quasiparticle band on the right for upper branch and other way round for the lower branch.
It is also easy to see this in Fig.~\ref{sdtri} and the upper panel of Fig.~\ref{ncexp}.
As a consequence of that there is also intensive shadow FS sheets around ($\pi/a$,0) point on Fig.~\ref{sk_fs}
(upper right panel).
Rather intensive nondispersing states at about -1.0 eV within experimental data
 can be presumably associated with the lower Hubbard band and
possible admixture of some oxygen states.
Let us also suppose that high intensity at -0.3 eV for $X$ point may be interpreted not as a
van-Hove singularity of bare dispersion~\cite{NdLDA} but rather of high-energy pseudogap branch.

\subsection{Comparison with optical data}

\begin{figure}

\includegraphics[clip=true,angle=270,width=0.8\columnwidth]{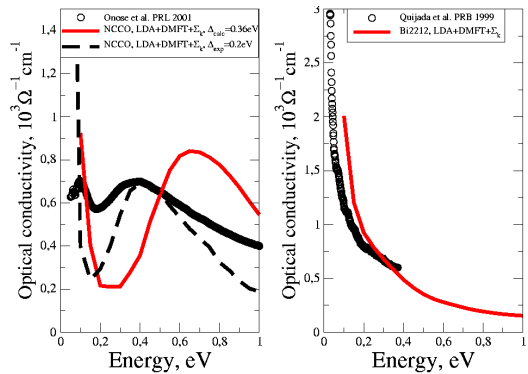}
\caption{Comparison of LDA+DMFT+$\Sigma_{\bf k}$ calculated optical conductivity spectra
for NCCO (left panel) with experimental data (circles) Ref.~\protect\cite{Onose}.
Solid line --- theoretical results for calculated pseudogap value $\Delta$=0.36 eV (dashed line corresponds to experimental $\Delta$=0.2 eV). On the right panel there is the same quantity but for Bi2212 and experiment of Ref.~\protect\cite{Quijada}.}
\label{opt}
\end{figure}

Our recent generalization of LDA+DMFT+$\Sigma_{\bf k}$ scheme with respect to two-particle properties~\cite{SkOpt}
allows us to analyze optical conductivity of the Bi and Nd materials under
consideration. In Fig.~\ref{opt} comparison of experimental data with part of optical conductivities for NCCO (left panel) and Bi2212 (right panel) is presented. 
The way of computation is described above.
Here we can report qualitative agreement of our theoretical curve for NCCO with calculated
$\Delta$=0.36 eV (solid line) with experiment~\cite{Onose}. Nevertheless
we find calculated pseudogap value to be overestimated.
To improve the agreement we also calculated optical conductivity for experimental value of $\Delta$=0.2~eV~\cite{Onose}
(Fig.~\ref{opt}, dashed line).
Concerning Bi2212 optical conductivity
(Fig.~\ref{opt}, right panel) one can point out that there is no particular structure neither in theory nor in the experimental data~\cite{Quijada}. For Bi2212 agreement between experimental and theoretical curves is reasonable.
Let us note that according to our calculations of the quasiparticle bands, spectral functions and FS maps
are not strongly modified in case of $\Delta$=0.2~eV.

\subsection{Influence of $U$ value on DMFT+$\Sigma_{\bf k}$ results}

Before we summarize our results,
we would like to discuss how the Hubbard interaction $U$ value affects 
DMFT+$\Sigma_{\bf k}$ results, namely, the observable physical quantities. 
This question arises from ongoing scientific discussion in the literature.
Note that our constrained LDA calculated $U$ value is of the order of 2-3$t$.
At the same time, it is commonly believed that $U$ value should be of the
order of 4-6$t$ \cite{Plakida,VT}. To this end we have performed  
additional DMFT+$\Sigma_{\bf k}$ computations, using these values of $U$.

If we take just higher values of $U$ within DMFT+$\Sigma_{\bf k}$ approach,
without any change of other parameters of our model,
we obtain stronger uniform quasiparticle damping. 
Spectral functions become slightly more blurred and FS less sharply defined. 
Also with the increase of the $U$ value quasiparticle mass grows up a little bit.
In optical conductivity the pseudogap anomaly also appears to be more damped, 
as we mentioned earlier in Ref.\cite{SkOpt}. At the same time the general
agreement with experiments stays reasonable.
This is not very surprising since these values still belong to small or
moderate correlations ($U$ is less than band width $W=8t$).

But actually, we are trying to express the pseudogap potential $\Delta$ 
via $U$ (see Appendix~\ref{app1}). If this connection is taken into account 
FS maps do not differ very much from those obtained above. However,
the pseudogap effects are getting stronger both around ``hot-spots'' and in 
the vicinity of $(\pi/2a,\pi/2a)$ {\bf k}-point.
In this case comparison with ARPES data for quasiparticle bands becomes much 
more worse. Especially, the bigger $U$ values spoil the accord with optical data.
Thus we can conclude, that constrained LDA calculated value of $U$ 
(together with corresponding value of $\Delta$) allows one to describe ARPES 
experimental data reasonably well, though in optical conductivity the size of 
the pseudogap is a bit overestimated. In other respects, we do not observe
any qualitative changes in our results as $U$ increase
from 3$t$ to 4$t$ or 6$t$.

\section{Conclusion}
\label{concl}

To summarize, the origin of evident ``hot-spots'' on NCCO FS in pseudogap 
regime is attributed to the details of noninteracting electronic band 
structure of this compound. All differences in physical properties calculated 
within the LDA+DMFT+$\Sigma_{\bf k}$ approach (quasiparticle bands, FS maps 
and ARPES data) are determined by the fact that in NCCO ``hot-spots'' lie 
closer to BZ center than in Bi2212. Also stronger AFM long range ordering 
tendency in NCCO favors the clearly visible ``hot-spots''.
Besides that, the qualitative behavior of both electron doped NCCO and hole 
doped Bi2212 high-T$_c$ systems is almost the same. 
We have also interpreted the new ARPES experimental data for NCCO
by LDA+DMFT+$\Sigma_{\bf k}$ calculated quasiparticle bands and proposed 
the new mechanism for the origin of van-Hove like singularity at -0.3~eV.

The results obtained further support the LDA+DMFT+$\Sigma{\bf k}$ to be an
effective method to investigate the electronic structure of strongly-correlated
systems.

\section{Acknowledgements}

We thank Thomas Pruschke for providing us the NRG code and O. Jepsen for helpful discussions.
This work is supported by RFBR grants 08-02-00021, 08-02-00712, 08-02-91200, RAS programs 
``Quantum macrophysics'' and ``Strongly correlated electrons in 
semiconductors, metals, superconductors and magnetic materials'', Dynasty 
Foundation, Grant of President of Russia MK-2242.2007.2, MK-3227.2008.2, SS-1929.2008.2, interdisciplinary 
UB-SB RAS project, Russian Science Support Foundation. IN thanks University of Osaka (Japan) for hospitality.

\appendix
\section{Particle-hole asymmetry of the pseudogap potential $\Delta$}
\label{app1}

\begin{figure}
\includegraphics[clip=true,angle=270,width=0.9\columnwidth]{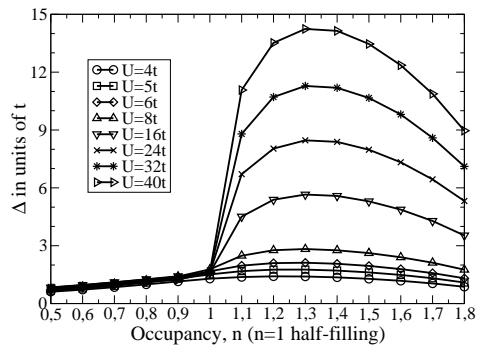}
\caption{Filling dependence of the pseudo gap potential $\Delta$
calculated with DMFT(QMC) and DMFT(NRG)
for varying Coulomb interaction ($U$) at temperature $T$=0.4$t$
on a two--dimensional square lattice with $t'/t$=-0.4.}
\label{delta}
\end{figure}

Pseudogap energy scale (amplitude) $\Delta$ was calculated in Ref.~\cite{cm05} 
via DMFT(QMC -- Quantum Monte Carlo) simulations for the hole doped region.
Here we present similar results for the electron doped case obtained with
DMFT(NRG - Numerical Renormalization Group). Using the two--particle selfconsistent approach of Ref.~\cite{VT},
with the approximations introduced in Refs.~\cite{Sch,KS}, for
standard Hubbard model one can derive following microscopic expression
for $\Delta$:
\begin{eqnarray}
\Delta^2=
U^2\frac{<n_{i\uparrow}n_{i\downarrow}>}{n^2}<(n_{i\uparrow}-n_{i\downarrow})^2>,
\label{DeltHubb}
\end{eqnarray}
where we take into account only scattering by antiferromagnetic spin fluctuations.
The different local quantities here, such as  density
$n$ and  double occupancy $<n_{i\uparrow}n_{i\downarrow}>$ etc.,
can easily be calculated within the standard DMFT \cite{georges96}. As impurity
solver we have used NRG.

In Fig.~\ref{delta} we show our results for $\Delta$ for both electron and hole 
dopings. One can immediately see a remarkable (up to one order of magnitude)
particle-hole asymmetry of the Eq.~(\ref{DeltHubb})
for large values of $U$. For values of $U$ below or equal to $8t$ (which 
corresponds to week or moderate coupling case) this $\Delta$ particle-hole 
asymmetry is about factor of two.
Basically in case of the Eq.~(\ref{DeltHubb}) it comes
from particle hole asymmetry of double occupancy $<n_{i\uparrow}n_{i\downarrow}>$ value
which is enhanced as $U$ value increases.
These results agree with experimental 
observation for pseudogap effects to be stronger for electron doped systems.

\end{document}